\documentclass[twocolumn,showpacs,preprintnumbers,amsmath,amssymb]{revtex4}

\usepackage{graphicx}	
\begin{document}


\title{Random Fan-Out State Induced by Site-Random Interlayer Couplings}


\author{Ryo Tamura}
\affiliation{Institute for Solid State Physics, University of Tokyo, 5-1-5 Kashiwa-no-ha, Kashiwa, Chiba 277-8581, Japan}
\author{Naoki Kawashima}
\affiliation{Institute for Solid State Physics, University of Tokyo, 5-1-5 Kashiwa-no-ha, Kashiwa, Chiba 277-8581, Japan}
\author{Takafumi Yamamoto}
\affiliation{Department of Energy and Hydrocarbon Chemistry, Graduate School of Engineering, Kyoto University, Nishikyouku, Kyoto 615-8510, Japan}
\author{Cedric Tassel}
\affiliation{Department of Energy and Hydrocarbon Chemistry, Graduate School of Engineering, Kyoto University, Nishikyouku, Kyoto 615-8510, Japan}
\author{Hiroshi Kageyama}
\affiliation{Department of Energy and Hydrocarbon Chemistry, Graduate School of Engineering, Kyoto University, Nishikyouku, Kyoto 615-8510, Japan}



\begin{abstract}
We study the low-temperature properties of a classical Heisenberg model with site-random interlayer couplings on the cubic lattice. 
This model is introduced as a simplified effective model of Sr(Fe$_{1-x}$Mn$_{x}$)O$_2$, which was recently synthesized.
In this material,
when $x=0.3$,
$(\pi\pi\pi)$ and $(\pi\pi0)$ mixed ordering is observed by neutron diffraction measurements.
By Monte Carlo simulations,
we find an exotic bulk spin structure that explains the experimentally obtained results.
We name this spin structure the ``random fan-out state''.
The mean-field calculations provide an intuitive understanding of this phase being induced by the site-random interlayer couplings.
Since Rietveld analysis assuming the random fan-out state agrees well with the neutron diffraction pattern of Sr(Fe$_{0.7}$Mn$_{0.3}$)O$_2$,
we conclude that the random fan-out state is reasonable for the spin-ordering pattern of Sr(Fe$_{0.7}$Mn$_{0.3}$)O$_2$ at the low-temperature phase.
\end{abstract}


\maketitle

\section{Introduction}

Disorder in materials induces the frustration effect and makes it difficult for cooperative phenomena to occur\cite{Binder1986,Ryan1992,Fernandez-Baca1995,Gaulin2005,Diep2005,Lacroix2011}.
For example, in random magnets,
conventional magnetic orders such as ferromagnetic (FM) and N\'eel orders are suppressed by the frustration that arises from disorder.
Consequently,
the ordering temperature of random systems is lower than that of pure systems and the ordered phase disappears due to random interactions between magnetic moments in some cases.
Systems with disorder also have many features that are not observed in pure systems.
In some random magnetic systems,
interesting phases and transitions have been observed,
for example, spin glass\cite{Cannella1972,Mezard1987,Fischer1993,Young1998}, mixed ordering\cite{Wong1985,Matsubara1996,Nielsen1996,Ryan2003,Bekhechi2004,Beath2006}, and oblique phase\cite{Katsumata1979,Katsumata1983,Matsubara1977}.
To find phases having exotic static and dynamic behavior,
synthesis and characterization of random magnets have been attracted attention in materials science for a long time.

From the viewpoint of statistical physics, 
it is important to explore how disorder in magnetic systems affects phase transitions and magnetic orders. 
A theoretical model of random magnets constructed by random substitution of magnetic ions is a site-random model\cite{Matsubara1996,Nielsen1996,Bekhechi2004,Beath2006,Matsubara1974,Matsubara1975,Aharony1975,Fishman1979,Oguchi1979,Shirakura1993,Hukushima1997}. 
Regarding a binary magnetic alloy, 
we can describe a site-random model with two types of magnetic ions, A and B. 
In this model, 
A and B ions are placed randomly on the lattice. 
The interactions of each bond depend on the combination of ions. 
For example, 
the interaction between A ions is antiferromagnetic (AF), 
whereas other interactions are FM. 
The Heisenberg model using this decision rule of interactions for all directions on a three-dimensional lattice has been investigated by Monte Carlo simulations\cite{Matsubara1996,Nielsen1996,Bekhechi2004,Beath2006}. 
This model can be considered as a fundamental model of isotropic materials. 
However, 
we cannot adapt this model in the present case because origins of interaction depend on the direction in the infinite-layer structure treated in this paper.
In this case, 
the emergence of disorder of interaction should depend on the direction. 
In isotropic random systems, 
the key element is the concentration density of each ion. 
In anisotropic random systems, 
in addition to that, 
the spatial distribution of disorder is also important. 
Then it is an interesting issue to investigate phase transition and spin structure in anisotropic random systems. 
Since recent developments in synthetic methodology enable us to design the spatial structure of materials, 
the above-mentioned issue has become increasingly important in materials science.

\begin{figure}[t]
\includegraphics[trim=0mm 140mm 0mm 0mm ,scale=0.45]{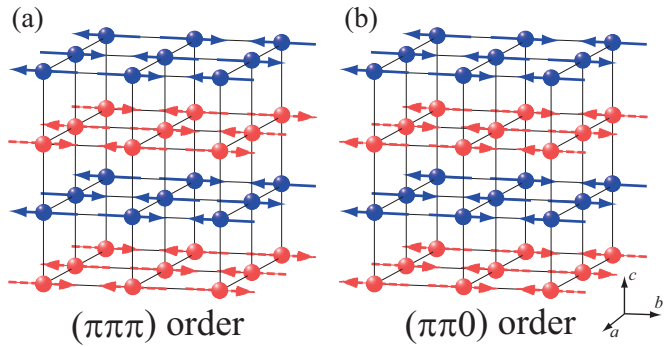} 
\caption{\label{fig:cg_configuration} 
(Color online) (a) Schematic spin structure of ($\pi\pi\pi$) order.
(b) Schematic spin structure of ($\pi\pi$0) order.
}
\end{figure}

Recently,
the infinite-layer iron oxide SrFeO$_2$ was synthesized by hydride reduction of SrFeO$_3$, which has the perovskite structure\cite{Tsujimoto2007}.
From the crystal structure of SrFeO$_2$, 
the nearest-neighbor spin interaction $J$ in a plane is due to a Fe-O-Fe superexchange interaction, 
and the interlayer interaction $J'$ is due to direct through-space overlap between Fe ions. 
First-principle calculations\cite{Xiang2008,Pruneda2008}  and inelastic neutron scattering measurements\cite{Tomiyasu2010} consistently revealed 
that both interactions are AF with $J \sim 3$ meV and $J' \sim 1$ meV. 
Owing to these strong AF interactions, 
SrFeO$_2$ exhibits ($\pi\pi\pi$) order (see Fig.~\ref{fig:cg_configuration} (a)) with a relatively high N\'eel temperature $T_\text{N}$ = 473 K. 
Seinberg {\it et al}. studied the effect of substituting Mn$^{2+}$ ions ($d^5$) for Fe$^{2+}$ ions ($d^6$) at room temperature (RT)\cite{Seinberg2011}.
RT neutron diffraction and M\"ossbauer spectroscopy studies on Sr(Fe$_{1-x}$Mn$_{x}$)O$_2$ ($x=0.0$, 0.1, 0.2, 0.3) showed 
that Mn substitution at Fe sites substantially destabilizes ($\pi\pi\pi$) order with increasing $x$, 
and Sr(Fe$_{0.7}$Mn$_{0.3}$)O$_2$ becomes paramagnetic at RT\cite{Seinberg2011}. 
From the Goodenough-Kanamori rule\cite{Goodenough1955,Kanamori1959},
the Fe-O-Mn superexchange intralayer interaction must be AF.
Thus,
the destabilization of ($\pi\pi\pi$) order suggests that the Fe-Mn interlayer interaction is FM and competes with the AF Fe-Fe interlayer interaction $J'$\cite{Seinberg2011}.
From the above facts,
the material Sr(Fe$_{1-x}$Mn$_{x}$)O$_2$ is an anisotropic random system.
It is possible that an unusual phase transition and a spin structure that comes from interlayer random couplings emerge at low temperature.

This paper presents low-temperature neutron diffraction results on Sr(Fe$_{0.7}$Mn$_{0.3}$)O$_2$, as well as a theoretical interpretation of those results. 
We find that  two distinct wave vectors, namely, $(\pi\pi\pi)$ and $(\pi\pi0)$, develop simultaneously in Sr(Fe$_{0.7}$Mn$_{0.3}$)O$_2$ at low temperature.
The aims of our study are to clarify the spin-ordering pattern and to explain the mechanism whereby the $(\pi\pi\pi)$ and $(\pi\pi0)$ wave vectors coexist.
In Sec. II, we present the experimental results on the random magnet Sr(Fe$_{1-x}$Mn$_{x}$)O$_2$.  
In Sec. III,
we introduce a classical Heisenberg model with site-random interlayer couplings as a simplified effective model of Sr(Fe$_{1-x}$Mn$_{x}$)O$_2$. 
In Sec. IV, 
we show the Monte Carlo simulation results of finite-temperature properties of the model introduced in Sec. III.
Drawing the phase diagram of temperature versus Mn ion concentration, 
we find the mixed phase which is characterized by wave vectors $(\pi\pi\pi)$ and $(\pi\pi0)$.
We discuss the spin-ordering pattern of this phase and the universality class of each phase transition.
In Sec. V,
to consider the emergence mechanism of the mixed phase observed in Monte Carlo simulations,
we investigate the effect of random interlayer couplings by mean-field calculations.
In Sec. VI, 
we discuss whether the spin-ordering pattern is robust with respect to the details of the model,
such as decision rules of interactions and easy-plane anisotropy.
We also show the results of Rietveld analysis for Sr(Fe$_{0.7}$Mn$_{0.3}$)O$_2$.
Section VII summarizes the paper and our main conclusions.


\begin{figure}
\includegraphics[trim=-10mm 180mm 0mm 0mm ,scale=0.66]{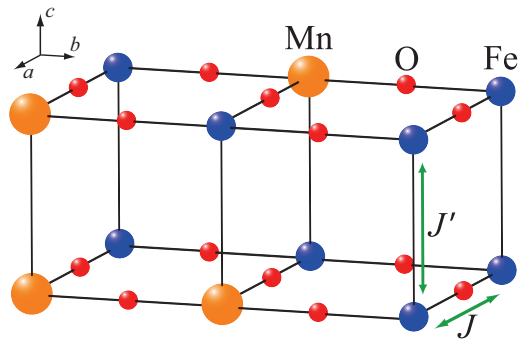} 
\caption{\label{fig:crystal} 
(Color online) (a) Crystal structure of Sr(Fe$_{1-x}$Mn$_{x}$)O$_2$.
Orange (large), blue (medium), and red (small) spheres indicate Mn, Fe, and O ions, respectively.
This arrangement of Fe and Mn ions is only an illustrative example.
In Sr(Fe$_{1-x}$Mn$_{x}$)O$_2$,
Fe and Mn ions are randomly placed on the tetragonal lattice.
}
\end{figure}

\section{Experimental Results}

The random magnet Sr(Fe$_{1-x}$Mn$_{x}$)O$_2$ ($x$ = 0.1, 0.2, 0.3) was recently synthesized\cite{Seinberg2011}. 
The crystal structure of Sr(Fe$_{1-x}$Mn$_{x}$)O$_2$ is shown in Fig.~\ref{fig:crystal}.
As we pointed out, the $x=0.1$ and 0.2 samples exhibit $(\pi\pi\pi)$ order above RT. 
The transition temperatures of the two substituted compounds have not been investigated,
but the experimental results in Ref.~\cite{Seinberg2011} indicate that $T_\text{N}$ decreases with increasing $x$.
Powder neutron diffraction experiments at RT show a systematic decrease in magnetic reflection intensity;
specifically, the relative intensity of the (1/2,1/2,1/2) reflection for $x$ = 0.0, 0.1, and 0.2 is 1:0.8:0.4.
$^{57}$Fe M\"ossbauer experiments at RT also reveal a systematic decrease of hyperfine fields:
402 kOe for $x$ = 0.0, 330 kOe for $x$ = 0.1, and 26.1 kOe for $x$ = 0.2.
When $x=0.3$, 
the system is in a disordered state at RT. 
Thus, 
we performed neutron diffraction measurements at low temperature. 

Powder neutron diffraction experiments on Sr(Fe$_{0.7}$Mn$_{0.3}$)O$_2$ were carried out on a Kinken powder diffractometer with multicounters for HERMES (High Efficiency and high Resolution MEasurementS) at the Institute for Materials Research, Tohoku University, installed at a guide hall of the JRR-3 reactor of the Japan Atomic Energy Agency (JAEA), Tokai\cite{Ohyama1998}.
The incident neutrons were monochromatized to 1.8204 \AA \ by the 331 reflection of bent Ge crystal. 
A 12'-blank-sample-18' collimation was employed. 
A polycrystalline sample of 3 g was placed into an He-filled vanadium cylinder. 
The sample temperature was controlled from 10 K to 273 K. 
The data were collected with a step-scan procedure using 150 neutron detectors in a 2$\theta$ range from 3$^\circ$ to 153$^\circ$ with a step width of 0.1$^\circ$. 

\begin{figure}
\includegraphics[trim=0mm 0mm 0mm 0mm ,scale=0.95]{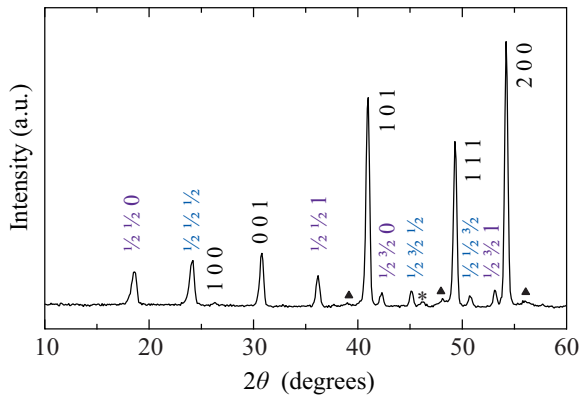} 
\caption{\label{fig:intensity} 
(Color online) Powder neutron diffraction pattern of Sr(Fe$_{0.7}$Mn$_{0.3}$)O$_2$ at 10 K. 
Purple and blue indices correspond to magnetic reflections of ($\pi\pi0$) and ($\pi\pi\pi$) orders, respectively. 
Triangles correspond to peaks from perovskite Sr(Fe$_{0.7}$Mn$_{0.3}$)O$_{3-\delta}$, 
and asterisk corresponds to a peak from an unknown impurity.
}
\end{figure}

\begin{figure}
\includegraphics[trim=-10mm 0mm 0mm 0mm ,scale=0.8]{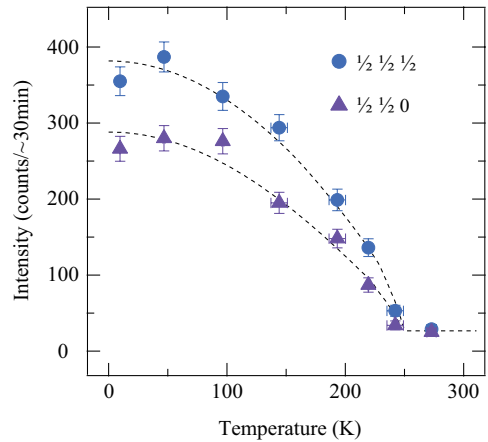}
\caption{\label{fig:intensity_temp} 
(Color online) Temperature dependences of (1/2,1/2,1/2) and (1/2,1/2,0) peaks corresponding, respectively, to ($\pi\pi\pi$) and ($\pi\pi0$) wave vectors 
for Sr(Fe$_{0.7}$Mn$_{0.3}$)O$_2$.
Dotted curves are visual guides. 
The intensities are normalized by acquisition time.
}
\end{figure}

Figure~\ref{fig:intensity} shows the neutron diffraction pattern for Sr(Fe$_{0.7}$Mn$_{0.3}$)O$_2$ at 10 K. 
As expected from the previous report\cite{Seinberg2011}, 
the diffraction profile has magnetic peaks such as (1/2,1/2,1/2) and (1/2,3/2,1/2) characterized by a wave vector ($\pi\pi\pi$), 
together with a nuclear reflection with a tetragonal $P4/mmm$ cell ($a$ = 4.001 \AA, $c$ = 3.441 \AA).
However, 
we unexpectedly observed magnetic peaks such as (1/2,1/2,0) and (1/2,1/2,1) characterized by a wave vector ($\pi\pi$0) at the same time. 
Plots versus temperature of the (1/2,1/2,1/2) and (1/2,1/2,0) peaks corresponding, respectively, 
to the ($\pi\pi\pi$) and ($\pi\pi$0) wave vectors are shown in Fig.~\ref{fig:intensity_temp}. 
Here, 
it is seen that the two peaks lose their intensities gradual with increasing temperature, 
and at and above 240 K, 
both peaks disappear and the intensities decrease to the background level. 
The pure SrFeO$_2$ system exhibits ($\pi\pi\pi$) order only, 
and thus the appearance of the ($\pi\pi$0) peak is caused by the effect of substitution on Mn ions.
We cannot predict the sign of direct exchange interaction along the interlayer direction,
because the direct exchange interaction strongly depends on the environment around it.
In contrast to the interlayer interaction,
from the Goodenough-Kanamori rule\cite{Goodenough1963},
Fe-O-Mn and Mn-O-Mn superexchange intralayer interactions must be antiferromagnetic.
Thus,
the simultaneous appearance of $(\pi\pi\pi)$ and $(\pi\pi0)$ wave vectors indicates that interlayer FM interactions appear at Mn sites.

A simple explanation of the observed simultaneous ($\pi\pi\pi$) and ($\pi\pi$0) wave vectors is phase separation into the ($\pi\pi\pi$)-type spin structure (Fig.~\ref{fig:cg_configuration} (a)) and the  ($\pi\pi$0)-type spin structure (Fig.~\ref{fig:cg_configuration} (b)). 
The other scenario is the emergence of a bulk spin structure in which ($\pi\pi\pi$) and ($\pi\pi0$) wave vectors coexist.
Since Sr(Fe$_{1-x}$Mn$_{x}$)O$_2$ is an anisotropic random system that has never been investigated theoretically,
it is possible this type of bulk spin structure emerged in Sr(Fe$_{1-x}$Mn$_{x}$)O$_2$.
Indeed, 
as we will see in the following theoretical argument, 
we find a novel type of magnetic structure that can explain the simultaneous appearance of ($\pi\pi\pi$) and ($\pi\pi$0) wave vectors. 
From these considerations, 
in Sec. VI B, 
we discuss the spin-ordering pattern in Sr(Fe$_{0.7}$Mn$_{0.3}$)O$_2$ at low temperature.

\section{Theoretical Model}
To investigate whether there is a bulk spin structure that explains the simultaneous appearance of ($\pi\pi\pi$) and ($\pi\pi0$) wave vectors without any phase separation,
we construct a simplified effective model of Sr(Fe$_{1-x}$Mn$_{x}$)O$_2$.
We introduce a classical Heisenberg model with site-random interlayer couplings on the cubic lattice.
This model includes the characteristic features of Sr(Fe$_{1-x}$Mn$_{x}$)O$_2$:
(i) the system consists of two types of magnetic ions randomly placed on the lattice and
(ii) competition between FM and AF interactions exists along only the interlayer direction.
The model Hamiltonian is given by
\begin{align}
\mathcal{H} &=J \sum_{\langle i,j \rangle_{\text{intralayer}}} \boldsymbol{s}_i \cdot \boldsymbol{s}_j + J' \sum_{\langle i,j \rangle_{\text{interlayer}}} \sigma_{ij} \boldsymbol{s}_i \cdot \boldsymbol{s}_j, \label{eq:hamiltonian}\\
\sigma_{ij} &:= - \frac{1}{2} \{ 1 + (\varepsilon_i + \varepsilon_j)-\varepsilon_i \varepsilon_j \},
\end{align}
where $\boldsymbol{s}_i$ is the three-component vector spin of unit length at the $i$-th site on the cubic lattice,
and $\varepsilon_i$ is equal to $-1$ or $+1$ for A or B ions at the $i$-th site, respectively.
Here,
A and B ions correspond to Fe and Mn ions in Sr(Fe$_{1-x}$Mn$_{x}$)O$_2$.
The first term of the Hamiltonian denotes the uniform AF nearest-neighbor interaction $(J>0)$ in the $ab$-plane.
The sign of the nearest-neighbor interaction in the second term depends on the arrangement of A and B ions along the interlayer direction ($c$-axis):
the interactions between A ions are assumed to be AF where $\sigma_{ij}=+1$,
and other interactions are assumed to be FM where $\sigma_{ij}=-1$.
We assume that absolute values of the interactions are the same ($J'=J$) for simplicity.
The interactions of this model are summarized in Table~\ref{tab:interaction}.
Under this rule of interactions,
we assume that the interlayer interaction at B ion sites is always FM.
However,
there are other possible decision rules of interactions where FM interactions exist at B ion sites.
In Sec. VI,
we discuss the low-temperature properties depending on the decision rule of interactions.

We calculate average densities of FM and AF interactions along the interlayer direction depending on a B ion concentration, $x$.
The probabilities of A-A, A-B, and B-B for nearest-neighbor ions along the interlayer direction are $(1-x)^2$, $2 x(1-x)$, and $x^2$, respectively.
Thus, 
probabilities of FM and AF interactions are $x^2+2x(1-x)$ and $(1-x)^2$, respectively.
The spatial average of signs of interlayer interactions, $\overline{\sigma}$, is given by
\begin{align}
\overline{\sigma}= 2 x^2 - 4 x +1,
\end{align}
where $\overline{\sigma}$ is equal to 0 at $x = x^* := (2-\sqrt{2})/2 \simeq 0.2929$.
In the case of $x<x^*$,
the AF interaction is dominant ($\overline{\sigma}>0$) along the interlayer direction.
In contrast,
the FM interaction is dominant ($\overline{\sigma}<0$) for $x>x^*$.

In the site-random model,
the signs of interactions along the interlayer direction are correlated,
and thus the FM correlation between next-nearest layers (NNLs) emerges.
In fact,
the effective FM interaction between NNLs exists regardless of $x$ in our model.
The sign of the effective interaction between next-nearest-neighbor sites is defined by 
\begin{align}
\sigma_{ik}^{\text{nnl}} := - \sigma_{ij} \sigma_{jk},
\end{align}
where $i$ and $k$ denote next-nearest-neighbor sites along the interlayer direction through the site $j$.
To show that effective interactions are always FM,
we investigate the arrangements of three ions including two interactions along the interlayer direction.
If the signs of both interactions are the same, that is, $\sigma_{ij}=\sigma_{jk}$,
the effective interaction is FM ($\sigma_{ik}^{\text{nnl}}=-1$).
In contrast,
if one of the interactions is FM coupling and the other is AF coupling, that is, $\sigma_{ij}=-\sigma_{jk}$,
the effective interaction is AF ($\sigma_{ik}^{\text{nnl}}=+1$).
For example,
the effective interactions of arrangements A-A-A and A-A-B are FM and AF, respectively.
We summarize the effective interactions and probabilities depending on arrangement of three ions in Table~\ref{tab:nnn_interaction}.
From Table~\ref{tab:nnn_interaction},
the probabilities of FM and AF effective interactions are $1-2x(1-x)^2$ and $2 x(1-x)^2$, respectively.
The spatial average of signs of effective interactions between NNLs, $\overline{\sigma^{\text{nnl}}}$, is given by
\begin{align}
\overline{\sigma^{\text{nnl}}} = 4x^3-8x^2+4x-1.
\end{align}
From this equation,
it is clear that $\overline{\sigma^{\text{nnl}}}$ is negative regardless of $x$ ($0 \le x \le 1$).
Thus,
the effective interaction between NNLs is always FM.
From this fact,
it follows that the spin arrangement between odd-numbered (even-numbered) layers is FM along the interlayer direction in ordered phases.
Note that the ferromagnetic correlation between NNLs exists in not only collinear spin structures but also non-collinear spin structures.
This is because thermal and disorder fluctuations cause the ferromagnetic effective interaction regardless of spin structure.

\begin{table}
\begin{center}
\caption{\label{tab:interaction}
Decision rule of interactions for combinations of ions ($J>0$).
}
\begin{tabular}{c|cc}
\hline\hline
 & \ intralayer \ & \ interlayer \ \\ \hline
A-A & $+J$(AF) & $+J$(AF) \\
A-B & $+J$(AF) & $-J$(FM) \\
B-B & $+J$(AF) & $-J$(FM) \\
\hline\hline
\end{tabular}
\end{center}
\end{table}

\begin{table}
\begin{center}
\caption{\label{tab:nnn_interaction}
Effective interactions and probabilities of next nearest-neighbor spin pairs along the interlayer direction depending on the arrangement of three ions. 
}
\begin{tabular}{c|cc}
\hline\hline
\ arrangement \ & \ effective interaction \ & \ probability \ \\ \hline
A-A-A & FM ($\sigma_{ik}^\text{nnl}=-1$) & $(1-x)^3$ \\
A-A-B & AF ($\sigma_{ik}^\text{nnl}=+1$) & $x(1-x)^2$ \\
A-B-A & FM ($\sigma_{ik}^\text{nnl}=-1$) & $x(1-x)^2$ \\
B-A-A & AF ($\sigma_{ik}^\text{nnl}=+1$) & $x(1-x)^2$ \\
A-B-B & FM ($\sigma_{ik}^\text{nnl}=-1$) & $x^2(1-x)$ \\
B-A-B & FM ($\sigma_{ik}^\text{nnl}=-1$) & $x^2(1-x)$ \\
B-B-A & FM ($\sigma_{ik}^\text{nnl}=-1$) & $x^2(1-x)$ \\
B-B-B & FM ($\sigma_{ik}^\text{nnl}=-1$) & $x^3$ \\
\hline\hline
\end{tabular}
\end{center}
\end{table}

To compare the properties of site-random model and bond-random (BR) model where FM and AF interactions are placed randomly on each bond,
we analyze the effective interactions for bond-random model in a similar way.
The spatial average of signs of interlayer interactions $\overline{\sigma_{\text{br}}}$ and that of effective interaction between NNLs 
$\overline{\sigma^{\text{nnl}}_{\text{br}}}$ are given by
\begin{align}
\overline{\sigma_{\text{br}}} &= 1-2y, \\
\overline{\sigma^{\text{nnl}}_{\text{br}}} &= 4y^2-4y+1,
\end{align}
where $y$ is the concentration of FM bonds along the interlayer direction, and $\overline{\sigma_{\text{br}}}$ is 0 at $y= y^* := 1/2$.
Since $\overline{\sigma^{\text{nnl}}_{\text{br}}}$ is equal to 0 at $y=y^*$,
the bond-random model does not have long-range order in the interlayer direction at $y=y^*$.
Thus,
the existence of the FM effective interaction between NNLs regardless of $x$ is a characteristic feature of the site-random model and is crucial for obtaining a finite transition temperature even at $x=x^*$ in our model.


\section{Simulation Results}

We study the finite temperature properties of the classical Heisenberg model with site-random interlayer couplings
defined by Eq.~(\ref{eq:hamiltonian}) on an $N=L \times L \times L$ simple cubic lattice with a periodic boundary condition.
We use Monte Carlo simulations in which the standard heat-bath method is adopted.
Before starting the simulations,
we specify the B ion concentration $x$ such that the number of B ions ($N_\text{B} := N \times x$) is an integer.
In each simulation,
we prepare random configurations of A and B ions and set the interaction of each bond according to Table~\ref{tab:interaction}.

\subsection{Phase diagram}

\begin{figure}
\includegraphics[trim=0mm 0mm 0mm 0mm ,scale=0.32, angle=270]{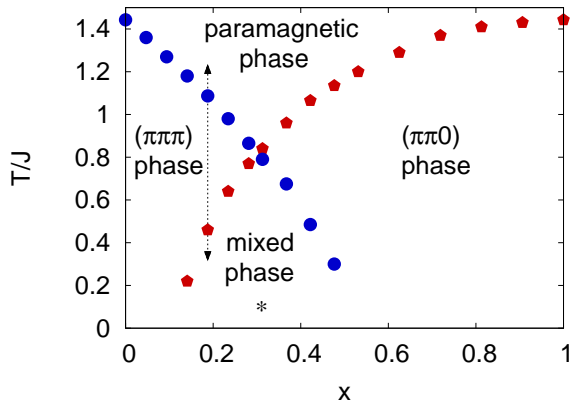} 
\caption{\label{fig:phase_diagram} 
(Color online) Phase diagram of temperature $T/J$ versus B ion concentration $x$ obtained from Monte Carlo simulations.
Asterisk and arrow denote parameters used in Sec. IV B (see Fig.~\ref{fig:correlation}) and C (see Fig.~\ref{fig:scaling}), respectively.
In the mixed phase,
$(\pi\pi\pi)$ and $(\pi\pi0)$ orders coexist,
and the spin-ordering pattern is the ``random fan-out state'' as shown in Fig.~\ref{fig:random_fanout}.
}
\end{figure}

In this section,
we draw the phase diagram of temperature $T/J$ versus B ion concentration $x$.
To determine transition temperatures,
we calculate the Binder ratio $U_4 (\boldsymbol{q})$ of magnetization vector $\boldsymbol{m} (\boldsymbol{q})$ characterized by wave vector $\boldsymbol{q}$:
\begin{align}
\boldsymbol{m} (\boldsymbol{q}) &:= \frac{1}{N} \sum_i \boldsymbol{s}_i e^{i \boldsymbol{q} \cdot \boldsymbol{r}_i}, \label{eq:mq} \\
U_4 (\boldsymbol{q}) &:= \left[ \frac{\left\langle \left| \boldsymbol{m} (\boldsymbol{q}) \right|^4 \right\rangle}{\left\langle \left| \boldsymbol{m} (\boldsymbol{q}) \right|^2 \right\rangle^2} \right]_\text{av}.
\end{align}
Here,
$\langle \cdots \rangle$ is the thermal average and $[\cdots]_\text{av}$ is the random average over arrangements of ions.
Hereafter,
$\boldsymbol{r}_i$ represents the position vector of the site $i$.
We prepare 64--1024 random configurations of ions.
The transition temperature is determined at the crossing point of $U_4 (\boldsymbol{q})$ by using $L=$ 12--32 data.
The phase diagram obtained from Monte Carlo simulations is shown in Fig.~\ref{fig:phase_diagram}.
There are three ordered phases in the phase diagram.
Here,
we determine each phase structure according to the structure factor:
\begin{align}
S(\boldsymbol{q}) := N \left[ \left\langle \left| \boldsymbol{m} (\boldsymbol{q}) \right|^2 \right\rangle \right]_\text{av}.
\end{align}
In the ($\pi\pi\pi$) and ($\pi\pi0$) ordered phases,
the structure factor has peaks only at each corresponding wave vector.
In the mixed phase,
structure factor peaks at two distinct wave vectors $(\pi\pi\pi)$ and $(\pi\pi0)$ develop.
There are two phase boundaries in the phase diagram.
The blue circles and red pentagons in Fig.~\ref{fig:phase_diagram} are determined by $U_4(\pi\pi\pi)$ and $U_4(\pi\pi0)$, respectively.

\subsection{Spin-ordering pattern in mixed phase}

\begin{figure}
\includegraphics[trim=0mm 0mm 0mm 0mm ,scale=0.32, angle=270]{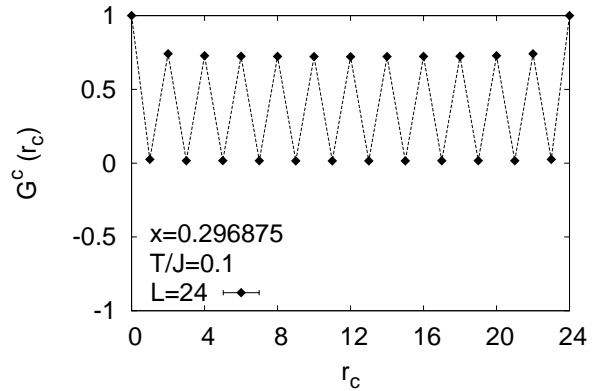} 
\caption{\label{fig:correlation} 
Correlation function between two spins along the interlayer direction ($c$-axis) for $L=24$ at $T/J=0.1$ obtained from Monte Carlo simulations.
Dotted line is visual guide.
}
\end{figure}

We next investigate the spin-ordering pattern in the mixed phase.
In the mixed phase,
peaks of structure factors at $(\pi\pi\pi)$ and $(\pi\pi0)$ develop;
namely,
N\'eel order appears in each layer.
To clarify the spin structure along the interlayer direction,
we calculate a correlation function between two spins along the interlayer direction ($c$-axis), which is defined by
\begin{align}
G^c(r_c) := \frac{1}{N} \left[ \sum_i \left\langle \boldsymbol{s} (\boldsymbol{r}_i) \cdot \boldsymbol{s} (\boldsymbol{r}_i+r_c \boldsymbol{e}_c)  \right\rangle \right]_\text{av}. \label{eq:correlation}
\end{align}
Here, 
$\boldsymbol{e}_c$ is a unit vector in the interlayer direction and $r_c$ is the distance between two spins.
Furthermore,
$\boldsymbol{s} (\boldsymbol{r}_i)$ denotes the spin at position $\boldsymbol{r}_i$, which is the same as $\boldsymbol{s}_i$.
We adopt $x=19/64=0.296875$, which is slightly larger than $x^*\simeq0.2929$.
Figure~\ref{fig:correlation} shows the dependence of distance $r_c$ of $G^c(r_c)$ for $L=24$ at $T/J=0.1$ which is below the transition temperature.
The spin correlation along the interlayer direction is FM for an even number of $r_c$.
This result is consistent with the expectation based on the effective interaction between NNLs discussed in Sec. III.
In contrast,
the spin correlation is nearly zero for an odd number of $r_z$.
This means that the angle between nearest-neighbor spin pairs along the interlayer direction is nearly $\pi/2$.
The value of spin correlation for an odd number of $r_c$ corresponds to the angle between spins belonging to odd-numbered and even-numbered layers.
Next,
we consider the $x$ dependence of this angle.

\begin{figure}
\includegraphics[trim=-10mm 0mm 0mm 0mm ,scale=0.4, angle=270]{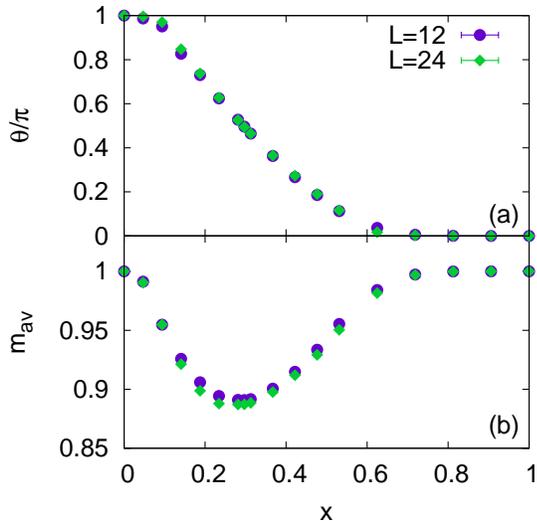} 
\caption{\label{fig:angle} 
(Color online) (a) Dependence on concentration $x$ of the angle $\theta/\pi$ in the ground state.
(b) Dependence on concentration $x$ of magnitude of magnetization $m_\text{av}:=(|\boldsymbol{m}_\text{o}|+|\boldsymbol{m}_\text{e}|)/2$ in the ground state.
}
\end{figure}

\begin{figure}
\begin{center}
\includegraphics[trim=0mm 170mm 0mm -10mm ,scale=0.48]{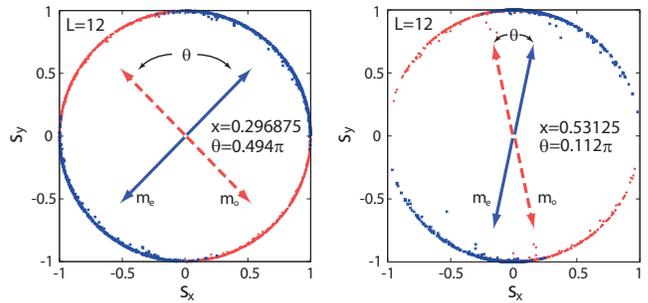} 
\caption{\label{fig:snapshot} 
(Color online) Snapshots of spin directions for $x=0.296875$ and 0.53125 when the lattice size is $L=12$ in the ground state.
Coordinates $(s_x, s_y)$ define the orthogonal plane of the vector $\boldsymbol{m}_\text{o} \times \boldsymbol{m}_\text{e}$.
}
\end{center}
\end{figure}

We calculate the $x$ dependence of the angle $\theta$ between the staggered magnetization vector of an odd-numbered layer $\boldsymbol{m}_\text{o}$ 
and that of an even-numbered layer $\boldsymbol{m}_\text{e}$ in the ground state.
The magnetization vectors $\boldsymbol{m}_\text{o}$, $\boldsymbol{m}_\text{e}$, and angle $\theta$ are defined by
\begin{align}
\boldsymbol{m}_\text{o} &:= \frac{2}{N} \sum_{\substack{ \boldsymbol{r}_i \in \text{odd-numbered} \\ \text{layer}}} e^{i \boldsymbol{q} \cdot \boldsymbol{r}_i} \boldsymbol{s}_i, \\
\boldsymbol{m}_\text{e} &:= \frac{2}{N} \sum_{\substack{ \boldsymbol{r}_i \in \text{even-numbered} \\ \text{layer}}} e^{i \boldsymbol{q} \cdot \boldsymbol{r}_i} \boldsymbol{s}_i, \\
\theta &:= \left[ \cos^{-1} \left\langle \frac{\boldsymbol{m}_\text{o} \cdot \boldsymbol{m}_\text{e}}{|\boldsymbol{m}_\text{o}||\boldsymbol{m}_\text{e}|} \right\rangle \right]_\text{av},
\end{align}
where $\boldsymbol{q} = (\pi\pi0)$.
We obtain the spin configuration in the ground state by the steepest descent method corresponding to zero temperature simulation.
We prepare 64 random configurations of ions and use the snapshot obtained from Monte Carlo simulations at $T/J=0.1$ as the initial spin configuration.
Figure~\ref{fig:angle} (a) shows the $x$ dependence of the angle $\theta/\pi$ in the ground state.
As the value of $x$ increases,
the value of $\theta/\pi$ changes from 1 corresponding to the ($\pi\pi\pi$) order to 0 corresponding to the ($\pi\pi0$) order.
Figure~\ref{fig:angle} (b) shows the $x$ dependence of $m_\text{av} := (|\boldsymbol{m}_\text{o}|+|\boldsymbol{m}_\text{e}|)/2$ in the ground state.
The value of $m_\text{av}$ is less than 1 in the mixed phase.
This means that the spin arrangement of each layer is not perfect N\'eel order.
Spin directions are distributed in a fan-shape around each magnetization axis owing to the effect of random fields from neighboring layers.
This fan-shaped distribution of spins can be seen in the snapshot of spin directions (Fig.~\ref{fig:snapshot}), which are projections onto the plane perpendicular to the vector $\boldsymbol{m}_\text{o} \times \boldsymbol{m}_\text{e}$.
Thus,
$|\boldsymbol{m}_\text{o}|$ and $|\boldsymbol{m}_\text{e}|$ indicate the shape of the spreading spin-fan.
Note that the absolute values of $\boldsymbol{m}_\text{o}$ and $\boldsymbol{m}_\text{e}$ should be the same, $|\boldsymbol{m}_\text{o}|\simeq|\boldsymbol{m}_\text{e}|$, 
because of the symmetry.
This fact is adopted in the mean-field analysis as an assumption, 
which is discussed in Sec. V.

\begin{figure}
\includegraphics[trim=0mm 165mm 0mm 0mm ,scale=0.45]{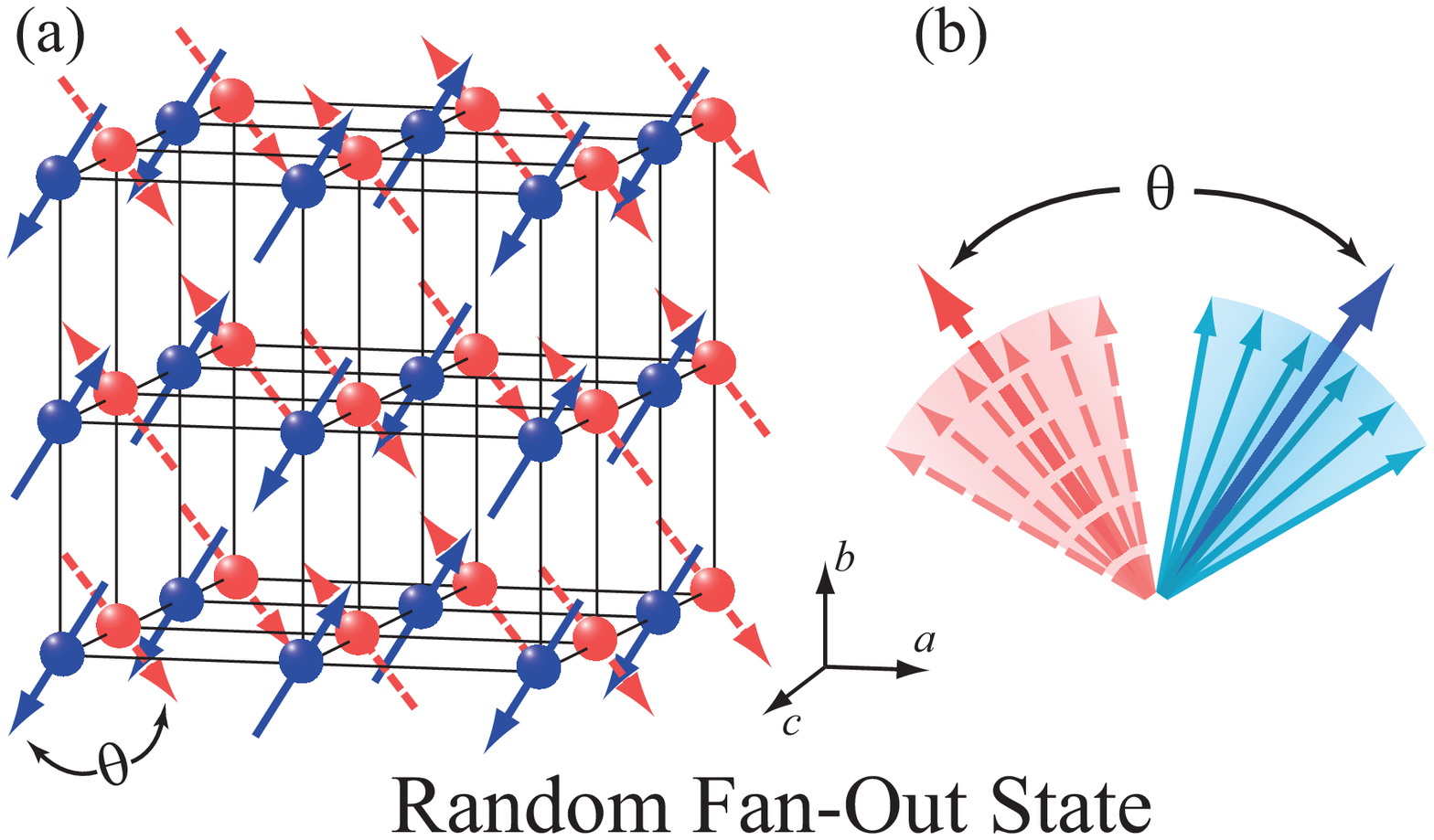} 
\caption{\label{fig:random_fanout} 
(Color online) (a) ``Average'' spin directions in the spin configuration of the random fan-out state.
In each layer ($ab$-plane),
N\'eel order appears.
Along the interlayer direction ($c$-axis),
the angle $\theta$ between nearest-neighbor ``average'' spin pairs changes from $\pi$ to 0 with increasing $x$ as shown in Fig.~\ref{fig:angle} (a).
The correlation between NNLs is FM.
(b) Individual spins are randomly directed around the average direction.
}
\end{figure}

\begin{figure*}
\includegraphics[trim=0mm 175mm 0mm 0mm ,scale=0.62]{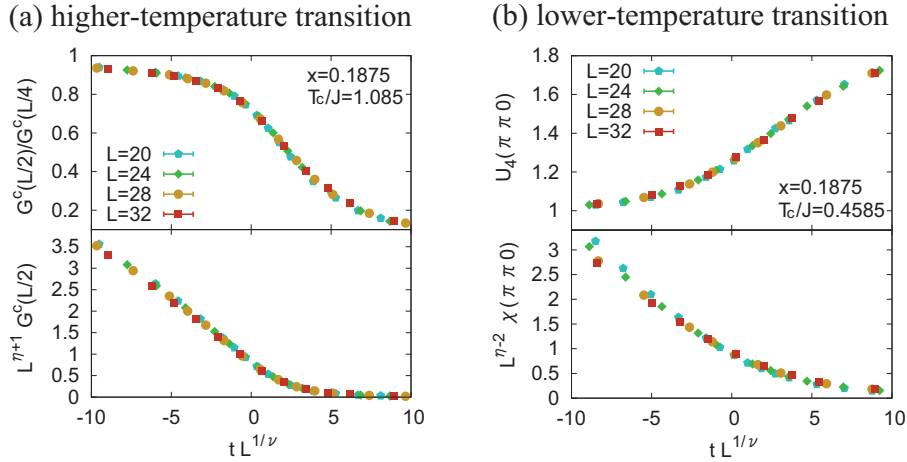} 
\caption{\label{fig:scaling} 
(Color online) (a) Finite-size scaling plots of the higher-temperature phase transition from the paramagnetic phase to the ($\pi\pi\pi$) ordered phase.
We adopt the critical exponents of three-dimensional Heisenberg universality class ($\nu=0.704$, $\eta=0.025$).
(b) Finite-size scaling plots of the lower-temperature phase transition from the ($\pi\pi\pi$) ordered phase to the mixed phase.
We adopt the critical exponents of three-dimensional XY universality class ($\nu=0.672$, $\eta=0.038$).
}
\end{figure*}

From these results,
the schematic spin configuration in the mixed phase can be depicted as shown in Fig.~\ref{fig:random_fanout}.
Figure~\ref{fig:random_fanout} (a) shows a schematic configuration of spin axes $\boldsymbol{m}_\text{o}$ and $\boldsymbol{m}_\text{e}$.
The direction of each arrow is the actual ``average'' magnetic direction.
In this configuration,
the spins are grouped into four sublattices according to whether they belong to an even/odd layer and which of the two sublattices in each layer.
Each sublattice corresponds to a distinct average spin direction as shown in Fig.~\ref{fig:random_fanout} (a).
Individual spins spread in the shape of fan around the average direction as shown in Fig.~\ref{fig:random_fanout} (b).
From these features,
we refer to this bulk spin-ordering pattern in the mixed phase as the ``random fan-out state''.
This random fan-out state is an exotic bulk spin ordering that explains the simultaneous appearance of ($\pi\pi\pi$) and ($\pi\pi0$) wave vectors without any phase separation.
In this model,
the ferromagnetic correlation between NNLs exists as explained in Sec. III.

\subsection{Universality classes of phase transitions}

We study the universality classes of phase transitions of our model.
In the phase diagram (Fig.~\ref{fig:phase_diagram}),
there are two types of phase boundaries.
To make clear the universality classes of each phase transition,
$x$ is set to $3/16=0.1875$ such that transition temperatures are separated sufficiently.
For this parameter,
the intermediate phase is the ($\pi\pi\pi$) ordered phase (see the dotted arrow in Fig.~\ref{fig:phase_diagram}).

First, we investigate the higher-temperature phase transition from the paramagnetic phase to the ($\pi\pi\pi$) ordered phase.
From the Harris criterion\cite{Harris1974},
we expect that the higher-temperature phase transition belongs to the three-dimensional Heisenberg universality class.
This is because the critical exponent $\alpha$ is negative in the three-dimensional Heisenberg model,
and thus the disorder should not affect the universality class.
To obtain the transition temperature and confirm the critical exponents,
we calculate the correlation function $G^c(r_c)$, which is defined by Eq.~(\ref{eq:correlation}).
The finite-size scaling relations of $d$-dimensional systems are given by
\begin{align}
\frac{G^c(L/2) }{G^c(L/4)} &\propto \Psi (t L^{1/\nu}), \\
G^c(L/2) &\propto L^{-d+2-\eta} \Phi (tL^{1/\nu}),
\end{align}
where $\Psi$ and $\Phi$ are scaling functions and $t := T-T_\text{c}$\cite{Tomita2002}.
We determine the transition temperature $T_\text{c}$ as the crossing point of $G^c(L/2)/G^c(L/4)$ using $L=$ 20--32 data and obtain $T_\text{c}/J=1.085(5)$.
This transition temperature is consistent with the one found for $U_4(\pi\pi\pi)$ in Sec. IV A.
The finite-size scaling using the critical exponents of the three-dimensional Heisenberg universality class ($\nu=0.704$, $\eta=0.025$)\cite{Chen1993}
are shown in Fig.~\ref{fig:scaling} (a).
Since the data are well fitted by scaling relations,
we conclude that the higher-temperature phase transition belongs to the three-dimensional Heisenberg universality class in accord with the Harris criterion.

Next,
we investigate the lower-temperature phase transition from the ($\pi\pi\pi$) ordered phase to the mixed phase.
The ($\pi\pi\pi$) ordered phase is translationally symmetric with the O(3) spin rotation symmetry broken down the spin rotation symmetry U(1).
In the mixed phase,
both the translational symmetry and the U(1) spin rotation symmetry are broken.
Thus,
we expect that the transition to the mixed phase is characterized by the breaking of U(1) spin rotation symmetry.
In other words,
we expect that the lower-temperature phase transition belongs to the three-dimensional XY universality class.
To obtain the transition temperature and confirm the critical exponents,
we calculate the magnetization vector $\boldsymbol{m} (\pi\pi0)$ defined by Eq.~(\ref{eq:mq}).
The finite-size scaling relations are given by
\begin{align}
U_4 (\boldsymbol{q}) &= \frac{\langle|\boldsymbol{m} (\boldsymbol{q})|^4\rangle}{\langle|\boldsymbol{m} (\boldsymbol{q})|^2 \rangle^2} \propto f (t L^{1/\nu}), \\
\chi (\boldsymbol{q}) &= N \frac{\langle |\boldsymbol{m} (\boldsymbol{q})|^2 \rangle}{T} \propto L^{2-\eta} g (tL^{1/\nu}),
\end{align}
where $f$ and $g$ are scaling functions, and $\boldsymbol{q} = (\pi\pi0)$.
We determine the transition temperature as the crossing point of $U_4(\pi\pi0)$ using $L=$ 20--32 data and obtain $T_\text{c}/J=0.4585(5)$.
The finite-size scaling using the critical exponents of the three-dimensional XY universality class ($\nu=0.672$, $\eta=0.038$)\cite{Hasenbusch1999}
are shown in Fig.~\ref{fig:scaling} (b).
Although we obtain a reasonably good fit,
it is not good enough to detect the small difference between XY critical exponents and the other critical exponents in a three-dimensional system.
However,
from the viewpoint of spin rotation symmetry,
we can deduce that the lower-temperature phase transition belongs to the three-dimensional XY universality class.


\section{Mean-field Calculations}

In this section,
to obtain an intuitive understanding of the emergence mechanism of the mixed phase,
we investigate the effect of random interlayer couplings by mean-field calculations.
For simplicity of notation,
we study the system where the intralayer interactions are FM.
Under the gauge transformation at alternating sites,
this model is equivalent to our model given by Eq.~(\ref{eq:hamiltonian}) in the case of no external field.
By applying the inverse of the gauge transformation to this model, 
we can obtain the same results as from the original model.
In this section,
we define $\boldsymbol{m}_\text{o}$ as the uniform magnetization vector of an odd-numbered layer and $\boldsymbol{m}_\text{e}$ as that of an even-numbered layer.
The magnitudes of these magnetization vectors are assumed to be the same ($m := |\boldsymbol{m}_\text{o}| = |\boldsymbol{m}_\text{e}|$),
which is a reasonable assumption as stated in Sec. IV B.
Here, 
let $\theta$ be the angle between $\boldsymbol{m}_\text{o}$ and $\boldsymbol{m}_\text{e}$.
In this mean-field calculation,
to consider a simpler explanation,
we focus on one odd-numbered layer with molecular fields from neighboring layers.
This is because the circumstances of odd-numbered layer and even-numbered layer are the same,
and thus it is enough to investigate either layer.
The molecular field from neighboring layers at the site $i$ in an odd-numbered layer is given by
\begin{align}
\boldsymbol{h}_i = -2J' \sigma_i \boldsymbol{m}_\text{e}, \label{eq:random_field}
\end{align}
where $J'$ $(>0)$ denotes the nearest-neighbor interaction along the interlayer direction, 
and $\sigma_i$ denotes the sign of interaction with neighboring interlayer sites.
Note that we have replaced the spins on the neighboring layers with the average spins $\boldsymbol{m}_\text{e}$ and neglected the fluctuation,
and the contributions of upper and lower layers are assumed to be the same.
Here,
we separate $\sigma_i$ into the spatial average $\overline{\sigma}$ and the deviation $\Delta \sigma_i$ from $\overline{\sigma}$ depending on the site 
($\sigma_i = \overline{\sigma}+\Delta \sigma_i$).
The random average of deviation $[\Delta \sigma_i]_\text{av}=0$ is assumed.
Accordingly,
$\boldsymbol{h}_i$ can be separated into
\begin{align}
\boldsymbol{h}_i &= \overline{\boldsymbol{h}}^\parallel + \Delta \boldsymbol{h}_i^{\parallel} + \Delta \boldsymbol{h}_i^{\perp},
\end{align}
where
\begin{align}
\overline{\boldsymbol{h}}^\parallel &:= - 2 J' m \cos \theta \overline{\sigma} \boldsymbol{e}_\text{e}^\parallel, \\
\Delta \boldsymbol{h}_i^\parallel &:= - 2 J' m \cos \theta \Delta \sigma_i \boldsymbol{e}_\text{e}^\parallel, \\
\Delta \boldsymbol{h}_i^\perp &:= - 2 J' m \sin \theta \Delta \sigma_i \boldsymbol{e}_\text{e}^\perp.
\end{align}
The symbols $\boldsymbol{e}_\text{e}^\parallel$ and $\boldsymbol{e}_\text{e}^\perp$, respectively, 
are unit vectors that are parallel and perpendicular direction of $\boldsymbol{m}_\text{e}$ to $\boldsymbol{m}_\text{o}$.
Here,
we did not include the uniform transverse component $\overline{\boldsymbol{h}}^\perp$ in the definition of $\boldsymbol{h}_i$,
because $\overline{\boldsymbol{h}}^\perp$ does not contribute to the free energy.

\begin{figure}
\includegraphics[trim=0mm 0mm 0mm 0mm ,scale=0.42, angle=270]{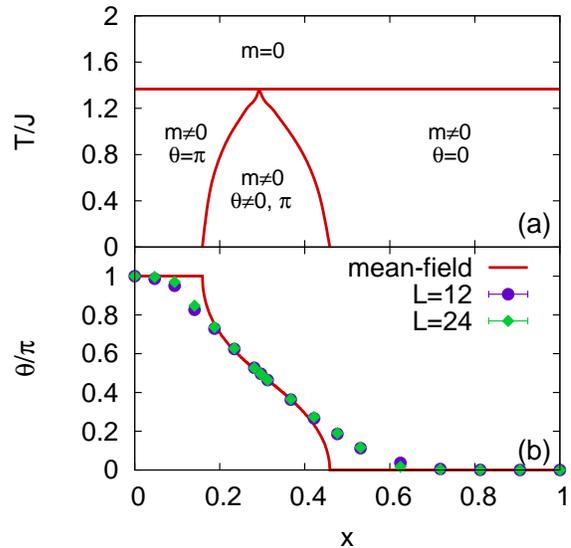} 
\caption{\label{fig:mean-field} 
(Color online)
(a) Phase diagram of temperature $T/J$ versus B ion concentration $x$ for $J=J'$ obtained from mean-field calculations.
(b) Dependence on concentration $x$ of angle $\theta/\pi$ between $\boldsymbol{m}_\text{o}$ and $\boldsymbol{m}_\text{e}$ in the ground state.
The points denote the simulation results for $L=12$ and 24, which are the same results shown in Fig.~\ref{fig:angle} (a).
}
\end{figure}

The excess free energy of a one-layer system from the effects of molecular fields $\boldsymbol{h}_i$ is given by
\begin{align}
\delta F(\theta) &= - \frac{1}{2} \sum_{i} \sum_{j} {}^t\!\boldsymbol{h}_i \chi_{ij} \boldsymbol{h}_j - \sum_{i} \overline{\boldsymbol{h}}^\parallel \cdot \boldsymbol{m}_i,
\end{align}
where $i$ and $j$ denote site indexes on one layer, and the susceptibility tensor is defined as
\begin{align}
\chi_{ij} = \frac{\partial \boldsymbol{m}_i}{\partial \boldsymbol{h}_j},
\end{align}
and $\boldsymbol{m}_i$ is the magnetization vector at the site $i$.
The random average of the product of deviations is expressed as
\begin{align}
[\Delta \sigma_i \Delta \sigma_j]_{\text{av}} &= (1 - \overline{\sigma}^2) \delta_{ij} =: \overline{\Delta \sigma^2} \delta_{ij}.
\end{align}
Hence, 
we obtain the free energy per site by taking the random average as following expression:
\begin{align}
\delta f &:= [\delta F(\theta)]_\text{av}/N \notag \\
&= - \frac{1}{2N} \Bigg\{ \sum_i \sum_j \chi_{ij}^\parallel \left( \overline{\boldsymbol{h}}^\parallel \right)^2 + \sum_i \chi_{ii}^\parallel \left[\left(\Delta \boldsymbol{h}_i^\parallel \right)^2 \right]_\text{av} \notag \\
&\ \ \ \ \ \ \ \ \ \ \ \ \ \ \ + \sum_i \chi_{ii}^\perp \left[ \left(\Delta \boldsymbol{h}_i^\perp \right)^2 \right]_\text{av} \Bigg\} - \frac{1}{N} \sum_i \overline{\boldsymbol{h}}^\parallel \cdot \boldsymbol{m}_i \notag \\
&= 2 J' m^2 \Bigg[ J' \Bigg\{ - \chi_{\text{uniform}}^{\parallel} \overline{\sigma}^2 \notag \\
&\ \ \ \ \ \ \ \ \ \ \ \ \ \ \ \ \ \ \ - \left( \chi_{\text{local}}^{\parallel} - \chi_{\text{local}}^{\perp} \right)  \overline{\Delta \sigma^2}  \Bigg\} \cos^2 \theta \notag \\
&\ \ \ \ \ \ \ \ \ \ \ \ \ \ \ \ \ \ \ \ \ \ \ \ \ \ \ \ + \overline{\sigma} \cos \theta - J' \chi_{\text{local}}^{\perp} \overline{\Delta \sigma^2} \Bigg], \label{eq:mean_free_energy}
\end{align}
where $N$ is the number of sites in one layer.
The symbols $\chi_\text{uniform}^\parallel$, $\chi_\text{local}^\parallel$, and $\chi_\text{local}^\perp$ represent the uniform longitudinal susceptibility,
the local longitudinal susceptibility, and the local transverse susceptibility, respectively:
$\chi_\text{uniform}^\parallel := \sum_{j} \chi_{ij}^\parallel$, $\chi_\text{local}^\parallel := \chi_{ii}^\parallel$, and $\chi_\text{local}^\perp := \chi_{ii}^\perp$ for all $i$.
In the mean-field approximation,
the magnitude of magnetization $m$ in the field $h$ along the magnetization vector is given by the Langevin function:
\begin{align}
L(\beta h) &= \coth(\beta h) - \frac{1}{\beta h} = m.
\end{align}
The molecular field from the same layer is given by
\begin{align}
h^\text{MF} = 4 J m.
\end{align}
Thus,
the susceptibilities obtained by mean-field approximation are calculated by using
\begin{align}
\chi_{\text{uniform}}^{\parallel} &= \frac{\chi_{\text{local}}^{\parallel}}{1- (h^{\text{MF}}/m) \chi_{\text{local}}^{\parallel}}, \\
\chi_{\text{local}}^{\parallel} &= \frac{\partial L(\beta h)}{\partial h} \Bigg{|}_{h=h^{\text{MF}}}, \\
\chi_{\text{local}}^{\perp} &= \frac{m}{h^{\text{MF}}}.
\end{align}
We calculate the angle $\theta$ that minimizes the excess free energy expressed by Eq.~(\ref{eq:mean_free_energy}).
The magnitude of the magnetization $m$ is obtained from the self-consistent equation:
\begin{align}
m = L \left(\beta h^{\text{MF}} \right).
\end{align}
The phase diagram of temperature $T/J$ versus B ion concentration $x$ for $J=J'$ is shown in Fig.~\ref{fig:mean-field} (a).
In this phase diagram,
there is a phase corresponding to the mixed phase where $\boldsymbol{m}_\text{o}$ and $\boldsymbol{m}_\text{e}$ are not parallel or antiparallel to each other.
We use the molecular field from the same layer only when we calculate the magnitude of magnetization $m$.
Then,
in the mean-field calculations,
the transition temperatures of the higher-temperature transition are clearly the same regardless of $x$.
To examine the properties of this phase,
we calculate the $x$ dependence of $\theta/\pi$ in the ground state (Fig.~\ref{fig:mean-field} (b)).
As the value of $x$ increases,
the value of $\theta/\pi$ changes from 1 corresponding to the $(\pi\pi\pi)$ order to 0 corresponding to the $(\pi\pi0)$ order.
Since the phase diagram and the behavior of $\theta$ obtained from mean-field calculations are qualitatively consistent with the Monte Carlo simulation results,
we expect that the emergence mechanism of the mixed phase is explained by mean-field calculations.

Next,
we consider the stabilization mechanism of the mixed phase.
From Eq.~(\ref{eq:mean_free_energy}),
when $\chi_\text{local}^\perp > \chi_\text{local}^\parallel$ and $(\Delta \boldsymbol{h}^\perp)^2 > 0$,
the excess free energy is minimized when $\theta/\pi \neq 1$ or 0.
Thus,
we conclude that the mixed phase emerges by the following mechanism.
(i) In the presence of finite magnetization,
spins are more susceptible to transverse field than longitudinal field, which is physically natural.
(ii) Transverse fields from neighboring layers exist as random fields $\Delta \boldsymbol{h}^\perp$ which come from the fluctuation of random interactions $\overline{\Delta \sigma^2}$.
This condition is realized by the existence of random interlayer couplings.
Thus,
we clarify that the mixed ordering, in which the magnetization vectors of neighboring layers are not parallel or antiparallel, 
is induced by random interlayer coupling.

\section{Discussion}

In this section,
we examine whether the random fan-out state is robust with respect to the decision rules of interactions and anisotropy. 
We also show the results of Rietveld analysis of Sr(Fe$_{0.7}$Mn$_{0.3}$)O$_2$ using the spin configuration obtained from Monte Carlo simulations at low temperature.

\subsection{Robustness of random fan-out state}

\begin{table*}[t]
\begin{center}
\caption{\label{tab:interaction_rule}
Decision rules of interactions along the interlayer direction.
Rule 1 is adopted in Sec. III, IV, and V.
}
\begin{tabular}{c|ccccccc}
\hline\hline
& $J_\text{AA}$ & $J_\text{AB}$ & $J_\text{BB}$ & $\overline{J}/J$ & $\overline{\Delta J}/J$ &$\overline{J^{\text{nnl}}}/J \propto$& $x^*$ \\ \hline
Rule 1 \ & \  $+J$(AF) \ & \ $-J$(FM) \ & \ $-J$(FM) \ & \ $2x^2-4x+1$ \ & \ $(1-\overline{J}^2)^{\frac{1}{2}}$ \ &$4 x (1-x)^2 -1$& \ $(2-\sqrt{2})/2$ \ \\
Rule 2 \ & \  $+J$(AF) \ & \ $-J$(FM) \ & \ 0 \ & \ $3x^2-4x+1$ \ & \ $(1-x^2-\overline{J}^2)^{\frac{1}{2}}$ \ &$3x^3-6x^2+4x-1$& \ $1/3$ \ \\
Rule 3 \ & \  $+J$(AF) \ & \ $+J$(AF) \ & \ $-J$(FM) \ & \ $-2x^2+1$ \ & \ $(1-\overline{J}^2)^{\frac{1}{2}}$ \ &$-4x^3+4x^2-1$& \ $1/\sqrt{2}$ \ \\
Rule 4 \ & \  $+J$(AF) \ & \ 0 \ & \ $-J$(FM) \ & \ $-2x+1$ \ & \ $(1-2x+2x^2-\overline{J}^2)^{\frac{1}{2}}$ \ &$-3x^2+3x-1$& \ $1/2$ \ \\
\hline\hline
\end{tabular}
\end{center}
\end{table*}

In this subsection,
we discuss the relationship between decision rules of interactions and the spin-ordering pattern.
In general,
the spatial average of interlayer random interactions $\overline{J}$ and its fluctuation $\overline{\Delta J}$ corresponding to the random fields depend on the signs and absolute values of interactions:
\begin{align}
\overline{J} &= J_\text{AA} (1-x)^2 + 2 J_\text{AB} x (1-x) + J_\text{BB} x^2, \\
\overline{\Delta J} &= \sqrt{J_\text{AA}^2 (1-x)^2 + 2 J_\text{AB}^2 x (1-x) + J_\text{BB}^2 x^2 - \overline{J}^2},
\end{align}
where $J_\text{AA}$, $J_\text{AB}$, and $J_\text{BB}$ denote the interactions between A-A, A-B, and B-B along the interlayer direction, respectively.
Then,
the spatial average of effective interaction between the NNLs, $\overline{J^{\text{nnl}}}$, is given by
\begin{align}
\overline{J^{\text{nnl}}} &\propto - J_\text{AA}^2 (1-x)^3 - 2 J_\text{AA} J_\text{AB} x (1-x)^2 \notag \\
&\ \  - J_\text{AB}^2 x (1-x)  - 2 J_\text{AB} J_\text{BB} x^2 (1-x) - J_\text{BB}^2 x^3.
\end{align}
Here,
we assume that $J_\text{AA}$ is always AF and the absolute values of interactions are the same or zero.
Then, 
four rules in Table~\ref{tab:interaction_rule} are all possible decision rules of interactions where FM and AF interactions exist and $\overline{J^{\text{nnl}}}<0$ regardless of $x$.
Note that rule 1 is adopted in the previous sections.
Under these rules and using Monte Carlo simulations and mean-field calculations,
we find the appearance of the mixed phase,
where the spin-ordering pattern is the random fan-out state.
Furthermore,
we also find that $(\pi\pi\pi)$ and $(\pi\pi0)$ magnetic peaks develop at the same temperature
and the angle between the staggered magnetization vectors of neighboring layers is $\pi/2$ at $x=x^*$ for all rules shown in Table~\ref{tab:interaction_rule}.

Next,
we consider the effect of easy-plane anisotropy.
We study the finite temperature properties of the XY spin version of our model by using Monte Carlo simulations.
Using rule 1 in Table~\ref{tab:interaction_rule},
we find the mixed phase where spin-ordering pattern is the random fan-out state.
Thus,
we conclude that the random fan-out state is not prohibited by the effects of easy-plane anisotropy.
In contrast,
the random fan-out state is prohibited by the effects of easy-axis anisotropy,
because the energy is raised when $\boldsymbol{m}_\text{o}$ and $\boldsymbol{m}_\text{e}$ are not parallel or antiparallel to each other.

From these discussions,
the random fan-out state is robust with respect to the details of our model, which satisfies the following conditions.
(i) The random FM-AF interactions exist along the interlayer direction.
(ii) The effective FM interaction between NNLs exists regardless of $x$.
(iii) There is no easy-axis anisotropy.

\subsection{Relation with experiments}

\begin{figure}[t]
\includegraphics[trim=40mm 0mm 0mm 0mm ,scale=1]{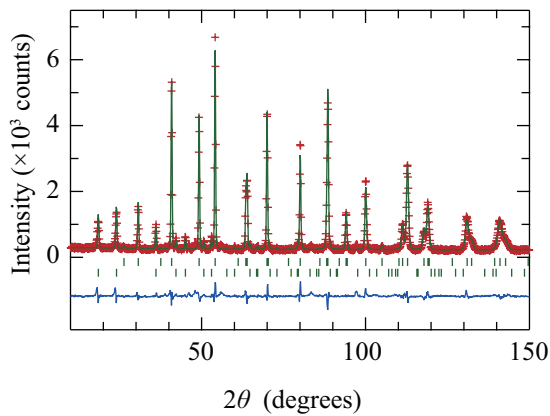} 
\caption{\label{fig:rietveld} 
(Color online) Observed (crosses), calculated (lines), and difference plots from the structural refinements of powder neutron diffraction data for Sr(Fe$_{0.7}$Mn$_{0.3}$)O$_2$ at 10 K.
The calculated data (lines) denote the intensity of the random fan-out state.
The upper and lower ticks represent the positions of the calculated chemical and magnetic reflections, respectively. 
}
\end{figure}

We discuss the spin structure in the low-temperature state of Sr(Fe$_{0.7}$Mn$_{0.3}$)O$_2$. 
In neutron diffraction measurements, 
magnetic peaks at ($\pi \pi \pi$) and ($\pi \pi 0$) develop at low temperatures as stated in Sec. II. 
This result can be explained by two possible scenarios---phase separation or a bulk spin structure that is the random fan-out state. 
In the former case, 
from the ratio of intensities at ($\pi \pi \pi$) and ($\pi \pi 0$), 
we can determine the volume fraction of each phase. 
In the latter case, 
the angle $\theta$ between the staggered magnetization vectors of neighboring layers can be estimated from the ratio of intensities at ($\pi \pi \pi$) and ($\pi \pi 0$). 
Therefore,
an important contrast between the two scenarios lies in the ordering temperature.
In the phase separation scenario, 
ordering of the two magnetic phases should occur at different temperatures because the absolute values of each interaction of Sr(Fe$_{0.7}$Mn$_{0.3}$)O$_2$ should be different.
On the other hand, 
in the bulk spin scenario, 
even when the absolute values of each interaction are different,
there should be a cross point of the phase boundary between the ($\pi \pi \pi$) and ($\pi \pi 0$) orders in the phase diagram. 
Based on experimental data,
Fig.~\ref{fig:intensity_temp} shows that the magnetic peaks at ($\pi \pi \pi$) and ($\pi \pi 0$) lose their intensities at nearly the same temperature,
which is a strong indication that the low-temperature state of Sr(Fe$_{0.7}$Mn$_{0.3}$)O$_2$ is likely to be the random fan-out state.
At the cross point,
the angle $\theta$ between the staggered magnetization vectors of neighboring layers should be nearly $\pi/2$ and the two magnetic peaks should disappear at nearly the same temperature. 
Thus, 
if the angle $\theta$ obtained by Rietveld structural refinement is nearly $\pi/2$, 
the simultaneous emergence of the two types of wave vectors can be naturally explained by the bulk spin structure scenario. 
To determine the angle $\theta$ in the random fan-out state, 
we perform Rietveld structural refinement of Sr(Fe$_{0.7}$Mn$_{0.3}$)O$_2$ using the random fan-out state. 
The results of Rietveld structural refinement of the neutron diffraction data and full details of the refined parameters are shown in Fig.~\ref{fig:rietveld} and Table~\ref{tab:rietveld_refinement}, respectively. 
The neutron diffraction pattern is well fitted by using the random fan-out state with $\theta=84^\circ$ in Fig.~\ref{fig:rietveld}. 
Thus, 
we conclude that it is highly possible that the mixed ordering in Sr(Fe$_{0.7}$Mn$_{0.3}$)O$_2$ is due to the bulk spin structure, 
which we call the random fan-out state.

Since the spin-ordering pattern in Sr(Fe$_{0.7}$Mn$_{0.3}$)O$_2$ at low temperature is explained by our model,
we expect that the $x-T$ phase diagram of Sr(Fe$_{1-x}$Mn$_{x}$)O$_2$ will agree with the theoretical phase diagram (Fig.~\ref{fig:phase_diagram}) qualitatively.
Indeed,
it appears that the transition temperature from the paramagnetic phase to the $(\pi\pi\pi)$ ordered phase $T_\text{N}$ decreases with increasing $x$ when $x=0.0$, 0.1, and 0.2 as stated in Sec. II.
This behavior of $T_\text{N}$ is consistent with the theoretical phase diagram (Fig.~\ref{fig:phase_diagram}).
If this theoretical phase diagram is valid for Sr(Fe$_{1-x}$Mn$_{x}$)O$_2$,
it will be observed that the $(\pi\pi0)$ peaks in Sr(Fe$_{1-x}$Mn$_{x}$)O$_2$ for $x=0.1$ and 0.2 develop at low temperatures.
In future work,
we will perform neutron diffraction measurements on Sr(Fe$_{1-x}$Mn$_{x}$)O$_2$ for several $x$ in addition to $x=0.3$ at low temperature.
Moreover,
we will study the relationship with the theoretical interpretation discussed in this paper.

\begin{table}
\begin{center}
\caption{\label{tab:rietveld_refinement}
Rietveld refinement for Sr(Fe$_{0.7}$Mn$_{0.3}$)O$_2^a$.
}
\begin{tabular}{lll}
\hline\hline
 & \ 10 K \ & \ RT$^b$ \ \\ \hline
$a$ (\AA) & 4.0014(1) \ \ & 4.0055(1) \\
$c$ (\AA) & 3.4406(1) & 3.4553(1) \\
$B$ in Sr (\AA) & 1.31(4) & 0.42(7) \\
$B$ in (Fe$_{0.7}$Mn$_{0.3}$) (\AA) \ \ & 1.08(4) & 0.21(6) \\
$B$ in O (\AA) & 1.51(5) & 0.91(5) \\
Moment ($\mu_{\text{B}}$) & 2.0 & 0 \\
Angle $\theta$ & 84$^\circ$ & 0 \\
$R_{\text{wp}}$ (\%) & 15.6 \% & 9.11 \% \\
$R_{\text{p}}$ (\%) & 15.2 \% & 7.01 \% \\
$\chi^2$ & 3.85 & 3.77 \\
\hline\hline
\end{tabular}
\end{center}
\begin{flushleft}
$^a$ All the refinements were performed by using the $P4/mmm$ space group with Sr on 1d (1/2,1/2,1/2), (Fe$_{0.7}$Mn$_{0.3}$) on 1a (0, 0, 0), and O on 2f (1/2, 0, 0). \\
$^b$ The refinement for RT is from Ref.~\cite{Seinberg2011}. 
\end{flushleft}
\end{table}


\section{Summary and Conclusion}

This paper is summarized as follows.
In Sec. II, 
we have reported the neutron diffraction pattern of Sr(Fe$_{0.7}$Mn$_{0.3}$)O$_2$.
We have observed that magnetic peaks corresponding to $(\pi\pi\pi)$ and $(\pi\pi0)$ wave vectors emerge simultaneously at $T_\text{N}=$ 240 K.

In Sec. III,
to investigate whether there is a bulk spin structure that explains the simultaneous appearance of $(\pi\pi\pi)$ and $(\pi\pi0)$ wave vectors,
we have introduced the classical Heisenberg model with site-random interlayer couplings on the cubic lattice
as a simplified effective model of Sr(Fe$_{1-x}$Mn$_x$)O$_2$.
This model consists of two types of magnetic ions, labeled A and B, that correspond to Fe and Mn ions.
The interactions in a plane have been assumed to be uniform antiferromagnetic.
The interactions along the interlayer direction between A ions have been assumed to be antiferromagnetic,
and other interactions have been assumed to be ferromagnetic.
In this decision rule of interactions,
the ferromagnetic effective interaction between next-nearest layers along the interlayer direction exists regardless of the B ion concentration $x$.
This is a characteristic feature of site-random interlayer couplings.

In Sec IV,
by performing Monte Carlo simulations,
we have shown the existence of a bulk spin structure that explains the simultaneous appearance of $(\pi\pi\pi)$ and $(\pi\pi0)$ wave vectors without any phase separation.
In this spin structure called the random fan-out state,
there are three features:
(i) the arrangement of spins in each layers is N\'eel order,
(ii) correlation between next-nearest layers has ferromagnetic,
(iii) the magnetization vector in odd-numbered layers and that in even-numbered layers are not parallel or antiparallel to each other.
As the value of $x$ increases,
the angle between these ordering vectors changes from $\pi$ corresponding to the $(\pi\pi\pi)$ order to 0 corresponding to the $(\pi\pi0)$ order.
Moreover,
we have constructed the phase diagram of temperature versus B ion concentration.
In the phase diagram,
the successive phase transitions have been observed.
We have also found the critical exponents of each phase transition.

In Sec. V, 
to clarify the emergence mechanism of the mixed phase,
we have investigated the effects of random interlayer couplings by using mean-field calculations.
It is important for the existence of a mixed phase that spins be more susceptible to transverse field than longitudinal field.
Since the transverse component of the random field from neighboring layers becomes large in the mixed phase,
the magnetization vectors of neighboring layers are not parallel or antiparallel.

In Sec. VI,
we have discussed whether the random fan-out state is robust with respect to the details of the model.
We have clarified that the random fan-out state becomes stable,
when random ferromagnetic-antiferromagnetic interactions exist and the ferromagnetic effective interaction between next-nearest layers exists regardless of $x$.
Furthermore,
we have determined that the random fan-out state is not prohibited by the effects of easy-plane anisotropy.
We have also performed Rietveld structural refinement of Sr(Fe$_{0.7}$Mn$_{0.3}$)O$_2$.
The neutron diffraction pattern is well fitted by using the random fan-out state.
Thus,
we conclude that the random fan-out state is a reasonable spin-ordering pattern of Sr(Fe$_{0.7}$Mn$_{0.3}$)O$_2$ at low temperature.

In this research, 
we have found a bulk spin structure called the random fan-out state that explains the simultaneous appearance of ($\pi\pi\pi$) and ($\pi\pi0$) wave vectors without any phase separation.
The model has been introduced as a simplified effective model of Sr(Fe$_{1-x}$Mn$_{x}$)O$_2$, 
and the neutron diffraction pattern of Sr(Fe$_{0.7}$Mn$_{0.3}$)O$_2$ is well fitted by using this spin structure. 
In our model, 
we have introduced the decision rule of interactions which induces disorder of interaction along only the interlayer direction. 
Thus, 
the spatial distribution of frustration in this model is different from that in systems having an isotropic crystal structure. 
Indeed, 
in a plane, 
since frustration does not exist, 
N\'eel order appears. 
In contrast, 
random ferromagnetic-antiferromagnetic interactions exist along the interlayer direction. 
Thus,
effect of frustration appears only between layers. 
At low temperature, 
the spin structure appears in which the angle between the neighboring layer magnetization axes are the same.
This structure is reminiscent of the spiral spin structure induced by the effects of frustration. 
However, 
in our site-random model, 
the interactions along the interlayer direction are correlated, 
and ferromagnetic correlation between next-nearest layers appears. 
Thus, 
a spiral spin structure characterized by a single wave vector does not appear, 
but an unusual spin-ordering pattern expressed by two wave vectors appears. 
Furthermore, 
in our model, 
the spin-fan structure exists in the ground state without an external field. 
In frustrated magnetic systems, 
the well-known spin-fan structure is realized by applying an external field along the direction parallel with the spiral plane\cite{Kitano1964,Komatsubara1970,Tachiki1970}. 
However, 
since the spin-fan structure of our model arises from an intrinsic effect of site-random couplings,
the emergence mechanism of the spin-fan structure is different from that of the conventional one\cite{Kitano1964,Komatsubara1970,Tachiki1970}. 
In our model, 
by the fluctuation of random interactions, 
the random fields from neighboring layers exist depending on the site, 
and thus spin directions have a fan-shaped distribution around each magnetization axis.

The random fan-out state is a novel type of spin-ordering pattern that comes from site-random interlayer couplings.
Thus, 
it is important to investigate the response to magnetic field or other external fields and spin wave excitation of this spin structure.
Recently, 
many exotic physical properties induced by competition and harmonization between spin structure and other degrees of freedom have been reported\cite{Tokura2006,Tanaka2006,Khomskii2009,Nagaosa2010,Ohkoshi2010,Chen2010}, 
and thus the search for new physical properties hidden in the random fan-out state is extremely important from the perspective of materials science.

In statistical physics, 
the phase transition and characteristic spin structure in anisotropic random systems will also be important areas of investigation. 
Although dynamical properties in random-spin systems have been studied for a long time, 
the relation between the dynamics and spatial distribution of disorder has not been established yet to the best of our knowledge,
and thus remains a challenging problem to be investigated.


\section*{Acknowledgment}

We thank Shu Tanaka, Takafumi Suzuki, Yusuke Tomita, Yuji Sumida, Liis Seinberg, Kazuyoshi Yoshimura, and Mikio Takano for useful comments and discussions.
R.T. was partially supported by the Global COE Program ``The Physical Sciences Frontier'', MEXT, Japan.
T.Y. was partially supported by the Japan Society for the Promotion of Science for Young Scientists. 
This research was supported by Grants-in-Aid for Scientific Research (B) (22340111) and for Scientific Research on Priority Areas ``Novel States of Matter Induced by Frustration'' (19052004) from MEXT, and by the Next Generation Supercomputing Project, Nanoscience Program, MEXT, Japan.
This work was also supported by the Japan Society for the Promotion of Science (JSPS) through its ``Funding Program for World-Leading Innovative R\&D on Science and Technology (FIRST) Program''.
The computations in the present work were performed on computers at the Supercomputer Center, Institute for Solid State Physics, University of Tokyo.


\end{document}